\documentstyle[aps,preprint,psfig]{revtex}

\begin{document}

\title{Emergence of Scaling in Random Networks}
\author{Albert-L\'aszl\'o Barab\'asi\thanks{alb@nd.edu} and R\'eka Albert}
\address{Department of Physics,
        University of Notre-Dame, Notre-Dame, IN 46556}
\maketitle
\vspace{2cm}

Systems as diverse as genetic networks or the world wide web are best described as networks with complex topology. A common property of many large networks is that the vertex connectivities follow a scale-free power-law distribution. This feature is found to be a consequence of the two generic mechanisms that networks expand continuously  by the addition of new vertices, and new vertices attach preferentially to already well connected sites. A model based on these two ingredients reproduces the observed stationary scale-free distributions, indicating that the development of large networks is governed by robust self-organizing phenomena that go beyond the particulars of the individual  systems.

\newpage

The inability of contemporary science to describe systems composed of non-identical elements that have diverse and nonlocal interactions currently limits advances in many disciplines, ranging from molecular biology  to computer science ({\it 1}). The difficulty in describing these systems lies partly in their topology: many of them form rather complex networks, whose vertices are the elements of the system and edges represent the interactions between them. For example, living systems form a huge genetic network, whose vertices are proteins and genes,  the edges representing the chemical interactions between them ({\it 2}). At a different organizational level, a large network is formed by the nervous system, whose vertices are the nerve cells, connected by axons ({\it 3}). But equally complex networks occur in social science, where vertices are individuals or organizations, and the edges characterize the social interaction between them ({\it 4}), or describe the world wide web (www), whose vertices are  HTML documents connected by links pointing  from one page to another ({\it 5, 6}). Due to their large size and the complexity of the interactions, the topology of these networks is largely unknown.

Traditionally, networks of complex topology have been described using the random graph theory of Erd\H{o}s and R\'enyi (ER) ({\it 7}), but in the absence of data on large networks the predictions of the  ER theory were rarely tested in the real world. However, driven by the computerization of data acquisition, such topological information is increasingly available, raising the possibility of  understanding the dynamical and topological stability of large networks.

In this paper we report on the existence of a  high degree of self-organization characterizing  the large scale properties of complex networks. Exploring several large databases describing the topology of large networks that span as diverse fields as the www or the citation patterns in science we show that, independent of the system and the identity of its constituents, the probability $P(k)$ that a vertex in the  network interacts with $k$ other vertices decays as a power-law, following $P(k)\sim k^{-\gamma}$.  This result indicates that large networks self-organize into a scale-free state, a feature unexpected by all existing random network models. To understand the origin of this scale invariance, we show that existing network models fail to incorporate growth and preferential attachment, two key features of real networks. Using a model incorporating these two ingredients, we show that they are responsible for the power-law scaling observed in real networks. Finally, we argue that these ingredients play an easily identifiable and important role in the formation of many complex systems, implying that our results are relevant to a large class of networks observed in Nature.

While there are a large number of systems that form complex networks,   detailed topological data is available only for a few. The collaboration graph of movie actors represents a well documented example of a social network. Each actor is represented by a vertex, two actors being connected if they were cast together in the same movie. The probability that an actor has $k$ links (characterizing his or her popularity) has a power-law tail for large $k$, following $P(k) \sim k^{-\gamma_{actor}}$, where $\gamma_{actor}=2.3\pm 0.1$ (Fig.$\,$1A). A more complex network with over $800$ million vertices  ({\it 8}) is the  www,  where a vertex is a document and the edges are the links pointing from one document to another. The topology of this graph determines the web's connectivity and, consequently, our effectiveness in locating information on the www ({\it 5}). Information about $P(k)$  can be obtained using robots ({\it 6}), indicating that the probability that $k$ documents point to a certain webpage follows a power-law,  with $\gamma_{www}=2.1\pm 0.1$ ( Fig.$\,$1B) ({\it 9}). A network whose topology reflects the historical patterns of urban and industrial development is the electrical powergrid of western US, the  vertices representing generators, transformers and substations, the edges  corresponding to the high voltage transmission lines between them ({\it 10}). Due to the relatively modest size of the network, containing only 4941 vertices, the scaling region is less prominent, but is nevertheless approximated with a power-law with an exponent $\gamma_{power}\simeq 4$ (Fig.$\,$1C). Finally, a rather large, complex network is formed by the citation patterns of the scientific publications, the vertices standing for papers published in refereed journals, the edges representing links to the articles cited in a paper. Recently Redner ({\it 11}) has shown that the probability that a paper is cited $k$ times (representing the connectivity of a paper within the network) follows a power-law with exponent $\gamma_{cite}=3$.

The  above examples ({\it 12}) demonstrate that many large random networks
share the common feature that the distribution of their local connectivity is free of scale, following a power-law for large $k$, with an exponent $\gamma$ between 2.1 and 4 which is unexpected within the framework  of the existing network models. The random graph model of ER ({\it 7}) assumes that we start with $N$ vertices, and connect each pair of vertices with probability $p$. In the model the probability that a vertex has $k$ edges follows a Poisson distribution $P(k)=e^{-\lambda}{\lambda}^k/k!$, where 
$\lambda=N\left(\begin{array}{c}N-1\\k\end{array}\right)p^k(1-p)^{N-1-k}$. In the small world model recently introduced by Watts and Strogatz (WS) ({\it 10}), $N$ vertices form a one-dimensional lattice, each vertex being connected to its two nearest and next-nearest neighbors.  With probability $p$ each edge is reconnected to a vertex chosen at random. The long-range connections generated by this process decrease the distance between the vertices, leading to a small-world phenomenon ({\it 13}), often referred to as six degrees of separation ({\it 14}). For $p=0$ the probability distribution of the connectivities is $P(k)=\delta(k-z)$, where $z$ is the coordination number in the lattice, while for finite $p$, $P(k)$ is still peaked around $z$, but it gets broader ({\it 15}). A common feature of the ER and WS models is that the probability of finding a highly connected vertex (that is, a large $k$) decreases exponentially with $k$, thus vertices with large connectivity are practically absent. In contrast, the power-law tail characterizing $P(k)$ for the studied networks indicates that highly connected  (large $k$) vertices have a large chance of occurring, dominating the connectivity.

There are two generic aspects of real networks that are not incorporated in these models. First, both models assume that we start with a fixed number ($N$) of vertices, that are then randomly connected (ER model), or reconnected (WS model), without modifying $N$. In contrast, most real world networks are open, they form by the continuous addition of new vertices to the system, thus the number of vertices, $N$, increases throughout the lifetime of the network. For example, the actor network grows by the addition of new actors to the system, the www grows exponentially in time by the addition of new web pages ({\it 8}), the research literature constantly grows by the publication of new papers. Consequently, a common feature of these systems is that the network continuously expands by the addition of new vertices that are connected to the vertices already present in the system.

Second, the random network models assume that the probability that two vertices are connected is random and uniform. In contrast, most real networks exhibit preferential connectivity. For example, a new actor is cast most likely in a supporting role, with more established, well known actors. Consequently, the probability that a new actor is cast with an established one is much higher than casting with other less known actors. Similarly, a newly created webpage will more likely include links to well known, popular documents with already high connectivity, or a new manuscript is more likely to cite a well known and thus much cited paper than its less cited and consequently less known peer. These examples indicate that the probability with which a new vertex connects to the existing vertices is not uniform, but there is a higher probability to be linked to a vertex that already has a large number of connections.

We next show that a model based on these two ingredients naturally leads to the observed scale invariant distribution. To incorporate the growing character of the network, starting with a small number ($m_0$) of vertices, at every timestep we add a new vertex with $m$($\leq m_0$) edges that link the new vertex to $m$ different vertices already present in the system. To incorporate preferential attachment, we assume that the probability $\Pi$ that a new vertex will be connected to vertex $i$ depends on the connectivity $k_i$ of that vertex, such that $\Pi(k_i)=k_i/\sum_j k_j$. After $t$ timesteps the model leads to a random network with $t+m_0$ vertices and $mt$ edges. This network evolves into a scale-invariant state with the probability that a vertex has $k$ edges following a power-law with an exponent $\gamma_{model}=2.9\pm 0.1$ (Fig.$\,$2A). As the power-law observed for real networks describes systems of rather different sizes at different stages of their development, it is expected that a correct model should provide a distribution whose main features are independent of time. Indeed, as Fig.$\,$2A demonstrates, $P(k)$ is independent of time (and, subsequently, independent of the system size $m_0+t$), indicating that despite its continuous growth, the system organizes itself into a scale-free stationary state.

The development of the power-law scaling in the model indicates that growth and preferential attachment play an important role in network development. To verify that both ingredients are necessary, we investigated two variants of the model.  Model A  keeps the growing character of the network, but preferential attachment is eliminated by assuming that a new vertex is connected with equal probability to any vertex in the system (that is, $\Pi(k)=const=1/(m_0+t-1)$). Such a model (Fig.$\,$2B) leads to $P(k)\sim \exp(-\beta k)$, indicating that the absence of preferential attachment eliminates the scale-free feature of the distribution. In model B we start with $N$ vertices and no edges. At each time step we randomly select a vertex and connect it with probability $\Pi(k_i)=k_i/\sum_j k_j$ to vertex $i$ in the system. While at early times the model exhibits power-law scaling, $P(k)$ is not stationary: since $N$ is constant, and the number of edges increases with time, after $T\simeq N^2$ timesteps the system reaches a  state in which all vertices are connected. The failure of models A and B indicates that both ingredients, namely growth and preferential attachment, are needed for the development of the stationary power-law distribution observed in Fig.$\,$1. 

Due to the preferential attachment, a vertex that acquired more connections than another one will increase its connectivity at a higher rate, thus an initial difference in the connectivity between two vertices will increase further as the network grows. The rate at which a vertex acquires edges is $\partial k_i/{\partial t}=k_i/{2t}$, which gives $k_i(t)=m(t/t_i)^{0.5}$, where $t_i$ is the time at which vertex $i$ was added to the system (see Fig.$\,$2C), a scaling property that could be directly tested once time-resolved data on network connectivity becomes available. Thus older (smaller $t_i$) vertices increase their connectivity at the expense of the younger (larger $t_i$) ones,  leading with time to some vertices that are highly connected, a "rich-gets-richer" phenomenon that can be easily detected in real networks. Furthermore, this property can be used to calculate $\gamma$ analytically. The probability that a vertex $i$ has a connectivity smaller than $k$, $P(k_i(t)<k)$, can be written as $P(t_i>m^2t/k^2)$. Assuming that we add the vertices at equal time intervals to the system, we obtain that $P(t_i>m^2t/k^2)=1-P(t_i\leq m^2t/k^2)=1-m^2t/k^2(t+m_0)$. The probability density $P(k)$ can be obtained from $P(k)=\partial P(k_i(t)<k)/\partial k$, which, at long times, leads to the stationary solution
$$P(k)=\frac{2m^2}{k^3},$$
giving $\gamma=3$, independent of $m$. While it reproduces the observed scale-free distribution, the proposed model cannot be expected to account for all aspects of the studied networks. For this we need to model these systems in more detail. For example, in the model we assumed linear preferential attachment, that is $\Pi(k)\sim k$. However, while in general $\Pi(k)$ could have an arbitrary nonlinear form $\Pi(k)\sim k^\alpha$, simulations indicate that scaling is present only for $\alpha=1$. Furthermore, the exponents obtained for the different networks are scattered between $2.1$ and $4$. However, it is easy to modify our model to account for exponents different from $\gamma=3$. For example, if we assume that a fraction $p$ of the links is directed, we obtain $\gamma(p)=3-p$, which is supported by numerical simulations ({\it 16}). Finally, some networks evolve not only by adding new vertices, but by adding (and sometimes removing) connections between established vertices. While these and other system-specific features could modify the exponent $\gamma$, our model offers the first successful mechanism accounting for the scale-invariant nature of real networks.

Growth and preferential attachment are mechanisms common to a number of complex systems, including business networks ({\it 17, 18}), social networks (describing individuals  or organizations), transportation networks ({\it 19}), etc. Consequently, we expect that the scale-invariant state,  observed in all systems for which detailed data  has been available to us, is a generic property of many complex networks, its applicability reaching  far beyond the quoted examples. A better description  of these systems would help in understanding other complex systems as well, for which so far less topological information is available, including  such important examples as genetic or signaling networks in biological systems. We often do not think of biological systems as open  or growing, since their features are genetically coded. However, possible
scale-free features of genetic and signaling networks could reflect the
evolutionary history dominated by growth and aggregation of different constituents, leading
from simple molecules to complex organisms.  With the fast advances in
mapping out genetic networks, answers to these questions might not be
too far. Similar mechanisms could explain the origin of the social and economic disparities governing competitive systems, since the scale-free inhomogeneities are the inevitable  consequence  of self-organization due to the local decisions made by the individual vertices, based on information that is biased towards the more visible (richer) vertices, irrespective of the nature and the origin of this visibility.

{\bf References and Notes}

1. R. Gallagher and T. Appenzeller, {\it Science} {\bf 284}, 79 (1999); R. F. Service, {\it Science} {\bf 284}, 80 (1999).

2. G. Weng, U. S. Bhalla, R. Iyengar, {\it Science} {\bf 284}, 92 (1999).

3. C. Koch and G. Laurent, {\it Science} {\bf 284}, 96 (1999).

4. S. Wasserman and K. Faust, {\it Social Network Analysis}, (Cambridge University Press, Cambridge, 1994).

5. Members of the {\it Clever} project, {\it Sci. Am} {\bf 280}, 54 (June 1999).

6. R. Albert, H. Jeong and A.-L. Barab\'asi, Nature {\bf 401}, 130 (1999), see also http://www.nd.edu/~networks.

7. P. Erd\H{o}s, and  A. R\'enyi, {\it Publ. Math. Inst. Hung. Acad. Sci} {\bf 5}, 17 (1960); B. Bollob\'as, {\it Random Graphs} (Academic Press, London, 1985).

8. S. Lawrence and C. L. Giles, {\it Science} {\bf 280}, 98 (1998); {\it Nature} {\bf 400}, 107 (1999).

9. Note that in addition to the distribution of incoming links, the www displays a number of other scale-free features, characterizing the organization of the webpages within a domain [B. A. Huberman and L. A. Adamic, Nature {\bf 401}, 131 (1999)], the distribution of searches [B. A. Huberman, P. L. T. Pirolli, J. E. Pitkow and R. J. Lukose, {\it Science} {\bf 280}, 95 (1998)], or the number of links per webpage ({\it 6}).

10. D. J. Watts and S. H. Strogatz, {\it Nature} {\bf 393}, 440 (1998).

11. S. Redner, {\it European Physical Journal B} {\bf 4}, 131 (1998).

12. We also studied the neural network of the worm Caenorhabditis elegans ({\it 3, 10}) and the benchmark diagram of a computer chip\\
(http://vlsicad.cs.ucla.edu/$\sim$cheese/ispd98.html). We find that $P(k)$ for both is consistent with power-law tails, despite the fact that for {\it C. elegans} the relatively small size of the system ($306$ vertices) limits severely the data quality, while for the wiring diagram of the chips vertices with over $200$ edges have been eliminated from the database.

13. S. Milgram, {\it Psychol. Today} {\bf 2}, 60 (1967); M. Kochen (ed.) {\it The Small World} (Ablex, Norwood, NJ, 1989).

14. J. Guare, {\it Six Degrees of Separation: A play} (Vintage Books, New York, 1990).

15. M. Barth\'el\'emy and L. A. N. Amaral, {\it Phys. Rev. Lett.} {\bf 82},   15 (1999).

16. Note that for most networks the connectivity $m$ of the newly added vertices is not constant. However, choosing $m$ randomly will not change the exponent $\gamma$ [Y. Tu, private communication].

17. W. B. Arthur, {\it Science} {\bf 284}, 107 (1999).

18. Note that preferential attachment was also used to model correlations between stock prices [L. A. N. Amaral and M. Barth\'el\'emy, private communication].

19. J. R. Banavar, A. Maritan and A. Rinaldo, {\it Nature} {\bf 399}, 130 (1999).

20. We thank D. J. Watts for providing the C. elegans and the power grid data, B. C. Tjaden for supplying the actor data, H. Jeong for collecting the data on the www and L. A. N. Amaral for helpful discussions. This work was partially supported by NSF Career Award DMR-9710998.

\begin{figure}[ht]
\hspace{1in}
\psfig{figure=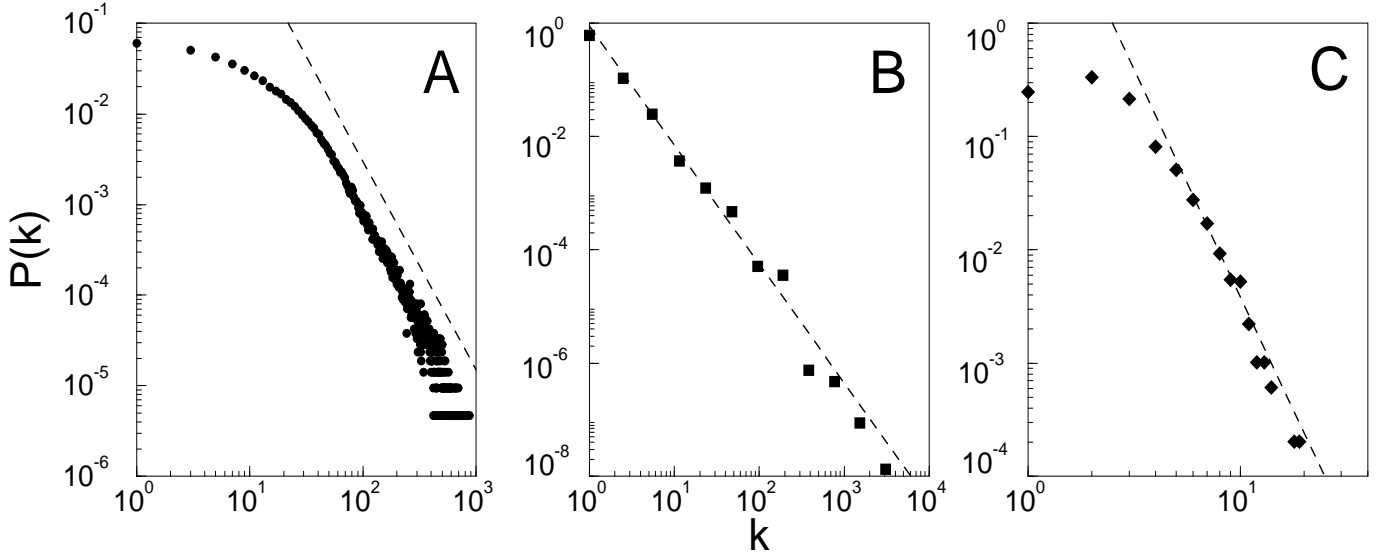,height=6.5in,width=2.6in,angle=-90}
\vspace{-2in}
\caption{The distribution function of connectivities for various large networks. ({\bf A}) Actor collaboration graph with $N=212,250$ vertices and average connectivity $\langle k\rangle =28.78$; ({\bf B}) World wide web, $N=325,729$, $\langle k\rangle=5.46$ ({\it 6}); ({\bf C}) Powergrid data, $N=4,941$, $\langle k \rangle=2.67$. The dashed lines have slopes ({\bf A}) $\gamma_{actor}=2.3$, ({\bf B}) $\gamma_{www}=2.1$ and ({\bf C}) $\gamma_{power}=4$.}
\end{figure}

\begin{figure}[ht]
\hspace{1in}
\psfig{figure=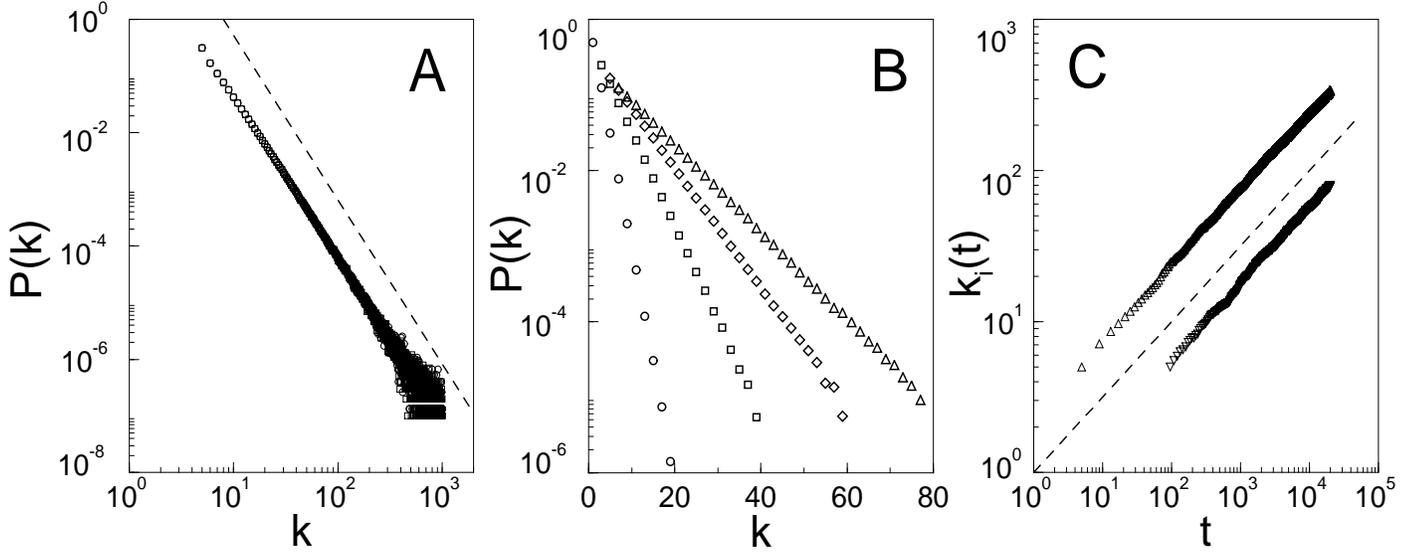,height=6.5in,width=2.2in,angle=-90}
\vspace{-2in}
\caption{({\bf A}) The power-law connectivity distribution at $t=150,000$ (o) and $t=200,000$ ($\Box$) as obtained from the model (see text), using $m_0=m=5$. The slope of the dashed line is $\gamma=2.9$. ({\bf B}) The exponential connectivity distribution for model A, in the case of $m_0=m=1$ (o), $m_0=m=3$ ($\Box$), $m_0=m=5$ ($\Diamond$) and $m_0=m=7$ ($\triangle$). ({\bf C}) Time evolution of the connectivity for two vertices added to the system at $t_1=5$ and $t_2=95$. The dashed line has slope $0.5$.}
\end{figure}

\end{document}